\documentclass[twocolumn,nofootinbib,superscriptaddress]{revtex4}

\usepackage{graphicx,graphics}
\usepackage{dcolumn}				
\usepackage{amssymb}

\usepackage{hyperref}
\usepackage{bm,pstricks}					

\newcommand{\beqa}{\begin{eqnarray}}
\newcommand{\eeqa}{\end{eqnarray}}
\newcommand{\beq}{\begin{equation}}
\newcommand{\eeq}{\end{equation}}

\newcommand{\env}{\mathcal{E}}
\newcommand{\sys}{\mathcal{S}}

\newcommand{\mean}[1]{\langle #1\rangle}

\begin{document}
\title{Entanglement dynamics during decoherence}

\author{Juan Pablo \surname{Paz}}
\affiliation{Departamento de F\'{\i}sica, FCEyN, UBA, Pabell\'on 1,
Ciudad Universitaria, 1428 Buenos Aires, Argentina}

\author{Augusto J. \surname{Roncaglia}}
\affiliation{Departamento de F\'{\i}sica, FCEyN, UBA, Pabell\'on 1,
Ciudad Universitaria, 1428 Buenos Aires, Argentina}

\begin{abstract}
The evolution of the entanglement between oscillators that interact with the same environment
displays highly non-trivial behavior in the long time regime. When the oscillators only interact 
through the environment, three dynamical phases were identified (J.P. Paz and A. Roncaglia, Phys. Rev. Lett. 100 (2008))
and a simple phase diagram characterizing them was presented. Here we generalize those results to the cases where 
the oscillators are directly coupled and we show how a degree of mixidness
can affect the final entanglement. In both cases, entanglement dynamics is fully characterized by three phases 
(SD: sudden death, NSD: no-sudden death and SDR: sudden death and revivals) which cover a phase diagram that 
is a simple variant of the previously introduced one. We present results when the oscillators are coupled to
the environment through their position and also for the case where the coupling is symmetric in position and
momentum (as obtained in the RWA). As a bonus, in the last case we present a very simple derivation of an 
exact master equation valid for arbitrary temperatures of the environment. 

\end{abstract}

\maketitle
\section{Introduction}

In recent years it became clear that the study of the evolution of entanglement for open quantum systems is an important
issue not only for fundamental reasons but also for practical ones, as entanglement is an essential resource for quantum
information processing \cite{Horodecki09}. Entanglement for systems of continuous variable is at the heart of the EPR 
argument \cite{Braunstein05} and was discussed in the context of recent experimental demonstrations of 
quantum teleportation \cite{Furu98} and cryptographic protocols \cite{Grosshans03}. The decoherence induced by the 
interaction with the environment is an important issue to consider in this context. In general, decoherence produces
dis-entanglement, which may occur in a finite time. This phenomenon \cite{Diosi03,Yu04,Almeida07} is known as
``sudden death'' of entanglement (SD). But the fate of entanglement for a quantum open system is not at all 
evident and many authors studied it obtaining rather surprising results for systems of qubits
\cite{Braun02,Kim02-2,Benatti03,Oh06,Anastopoulos06,Bellomo07} and for continuous 
variable systems \cite{Paris02,Serafini04,Olivares07,Dodd04,Prauzner04,Benatti06-2,Liu07,An07,Horhammer07,Zell09}. 

More recently, \cite{PazRoncaglia08,PazRoncaglia09}, we provided a unified picture that enabled us to understand the various
qualitatively different types of evolutions of the entanglement in non-Markovian environments. In fact, we showed that 
the asymptotic dynamics of entanglement can be described by three possible phases: 
SD (sudden death), SDR (sudden death and revival) and NSD (no sudden death). The fate of entanglement can be described 
using a simple phase diagram whose boundaries can even be analytical studied in some simple cases. In the above mentioned 
papers we presented the phase diagram under some simple assumptions: In particular, we assumed that the oscillators 
did not interact directly (only through the environment). Here, we briefly review the results 
of \cite{PazRoncaglia08,PazRoncaglia09} and we generalize them in two simple ways:  we consider more general initial
conditions and we also consider the case when the two oscillators directly interact between them. 

The paper is organized as follows. In Section \ref{sec:entangl} we review the basic notions of entanglement 
for two harmonic oscillators prepared in a Gaussian state. Here, we also describe the way in which the evolution of 
the entanglement can be studied and present a simple quantum optical analogy that enables us to understand the qualitative 
behavior displayed by the entanglement for long times. In Section \ref{sec:model} we discuss the models for system-environment 
coupling and present the phase diagram for the standard case briefly reviewing the results of \cite{PazRoncaglia08,PazRoncaglia09}. 
In Section \ref{sec:generalizations} we present the phase diagram for the case where there is a degree of impurity in the state 
of the virtual oscillators and direct interactions between the real oscillators. Finally, we summarize in Section \ref{sec:conc}. 
The Appendix \ref{ap:ME} contains a simple derivation of the exact master equation for an oscillator coupled with a bosonic 
environment through an interaction term which is symmetric under position and momentum interchange (similar to what is obtained 
under the usual RW approximation).

\section{Entanglement between two oscillators}
\label{sec:entangl}

We will consider a system of two identical quantum harmonic oscillators (with coordinates $x_1$ and $x_2$).
The interaction with the environment will be discussed in the next Section. Here we will discuss how can the 
entanglement between such oscillators be quantified and studied. We will restrict to consider Gaussian 
states (that will remain Gaussian under evolution according to the models described below) and factorized 
initial conditions between system and environment.
For such class of states we will be able to obtain simple analytical results. 

Entanglement for Gaussian states of two bosonic modes is entirely determined by the covariance matrix defined as 
\beq
V_{ij}(t)=\mean{\{r_i,r_j\}}/2-\mean{r_i}\mean{r_j},\nonumber
\eeq
where $i,j=1,\ldots,4$ and $\vec r=(x_1,p_1,x_2,p_2)$. In fact, a good measure of entanglement is the logarithmic 
negativity $E_{\mathcal N}$ \cite{Vidal02,PhdEisert,Adesso04} defined as: 
\beq
E_{\mathcal N}=\max\{0,-\ln(2\nu_{\min})\},
\eeq
where $\nu_{\rm min}$ is the smallest symplectic eigenvalue of the partially transposed covariance matrix. 
Therefore, the entanglement between the two oscillators is entirely determined by the second moments contained in $V_{ij}$. 
Thus, the strategy to study the evolution of the entanglement is obvious: We should find out the evolution of the elements 
of such matrix. In the following section we will show how to do this in a simple, but realistic, model. But it is useful to
advance here some of the most important results as they turn out to be rather independent of the details but only on the 
following important assumption: We will consider situations in which the system-environment interaction only involves a 
bilinear coupling in the collective coordinate $x_+=x_1+x_2$ (i.e., the relative coordinate $x_-=x_1-x_2$ will be 
effectively decoupled from the environment). In such cases, the asymptotic state will be such that the $x_+$ oscillator
will reach an equilibrium state characterized by the dispersions $\Delta x_+$ and $\Delta p_+$. Such dispersions depend
upon parameters of the model (spectral densities, initial temperature, etc) and such dependence will be discussed below.
For the moment we only need to assume the existence of an equilibrium state for $x_+$. Also, as equilibrium approaches, 
correlations between $x_+$ and $x_-$ vanish. On the other hand, the second moments of $x_-$ corresponds to a free oscillator
with a certain frequency $\omega_-$. 

The above simple observations are almost all we need to fully analyze the evolution of the entanglement between 
the two oscillators. Thus, in the long time limit, the covariance matrix has a simple block-diagonal from in terms 
of the variances of the $x_\pm$ oscillators. From this, it is possible to obtain covariances for the original $x_{1,2}$ oscillators 
and to find the smallest symplectic eigenvalue of the partially transposed version of such matrix. The result for 
the logarithmic negativity is: 
\beq
E_{\mathcal N}(t)\rightarrow\max\{0,E(t)\}, 
\eeq
where the function $E(t)$ is defined as
\beq
E(t)= \tilde E_{\mathcal N}+{\Delta E_{\mathcal N}} G(t). \label{eq:Eoft}
\eeq
Here $G(t)$ is an oscillatory function with period $\pi/\omega_-$ that takes values in the interval $\{ -1,+1 \}$.
The mean value $\tilde E_{\mathcal N}$ and the amplitude $\Delta E_{\mathcal N}$ that characterize the oscillations
of $E(t)$ are simply written as
\beqa
\tilde E_{\mathcal N}&=&\max\{|r|,|r_{crit}|\} -S_{crit},\label{eq:ent}
\\
\Delta E_{\mathcal N}&=&\min\{|r|,|r_{crit}|\}.
\label{eq:deltaE}
\eeqa
In the above equations $r$ is the initial squeezing factor defined in terms of the dispersions of the initial state 
$\delta x_-$ and $\delta p_-$ as 
\beq
r={1 \over 2}\ln\left[m_- \omega_- {\delta x_- \over \delta p_-}\right],\nonumber
\eeq 
and $r_{crit}$ is related to the squeezing of the equilibrium state for the  $x_+$-oscillator:
\beq
r_{crit}={1 \over 2}\ln\left[m_- \omega_- {\Delta x_+ \over \Delta p_+}\right]. 
\label{eq:rcrit}
\eeq 
Finally, $S_{crit}$ is defined as
\beq
S_{crit}={1\over 2} \ln[4 \Delta x_+ \Delta p_+ \delta x_- \delta p_-], 
\label{eq:Scrit}
\eeq
and turns out to be simply related with the entropy of the asymptotic state by the symplectic area of the oscillators $x_\pm$.

Using these results we can conclude that there are three qualitatively different types of evolutions for the entanglement 
for long times. First, entanglement may persists for arbitrary long times when $||r|-|r_{crit}||>S_{crit}$. In this case 
there is no sudden death of entanglement (NSD). A different behavior, an infinite sequence of events of sudden death and 
sudden revival (SDR), takes place when $||r|-|r_{crit}||<S_{crit}$ but $|r|+|r_{crit}|>S_{crit}$. Finally, a third  
phase characterized by a final event of sudden death (SD) of entanglement is realized if $|r|+|r_{crit}|<S_{crit}$.

\subsection{Interpretation: Where does the entanglement come from?}
\label{sec:interp}

\begin{figure}[htb!]
\includegraphics[width=8.7cm]{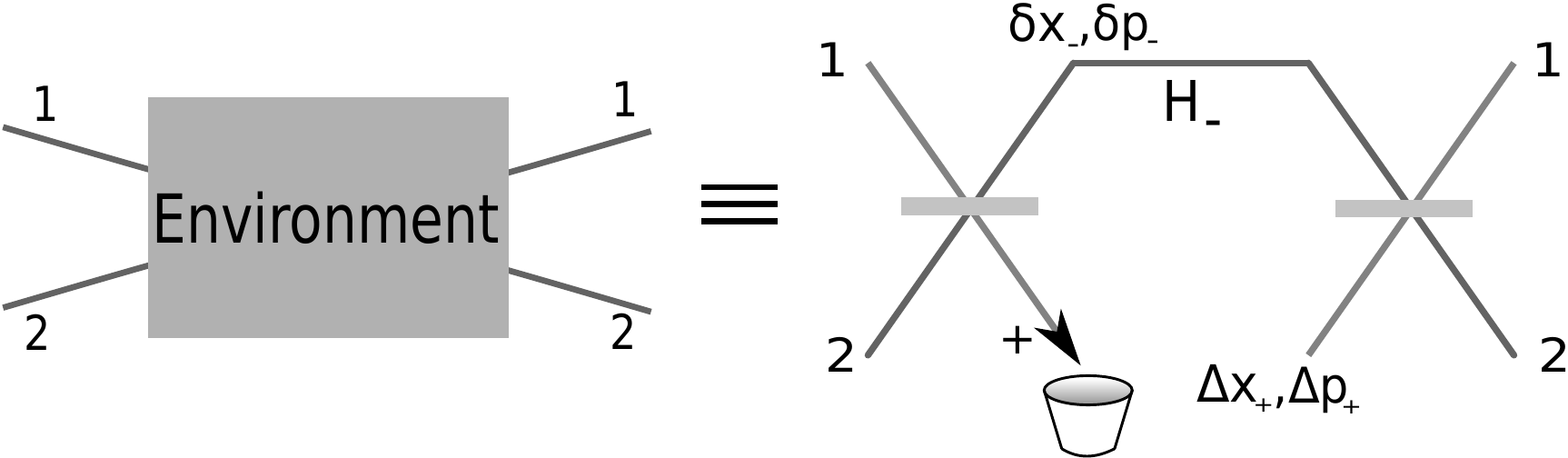}
\caption{The evolution of two resonant oscillators coupled to a common environment (left side of the diagram)
turns out to be described, in the long time regime, by the right side of diagram: A $50/50$ beam splitter 
combines the original $x_{1,2}$ oscillators to form the $x_{\pm}$ modes. While $x_-$ evolves freely, 
$x_+$ couples to the environment which drives this mode to an asymptotic equilibrium state which is 
completely uncorrelated with $x_-$ and is fully characterized by the dispersions $\Delta x_+$ and $\Delta p_+$.
A second beam splitter recreates the $x_{1,2}$ modes which will be entangled only if there is squeezing in 
the $x_{\pm}$ modes before the second beam splitter.
} 
\label{fig:interpret}
\end{figure}

The above results may seem, at first sight, somewhat puzzling. The final state of the two oscillators may be entangled even
if there was no entanglement in the initial state. In some sense, the common environment provides a quantum channel through
which entanglement between the two oscillators can be either created or destroyed, depending on the circumstances 
(initial state, temperature, etc). Here, we will discuss this result and present a very simple interpretation. 
The key is to realize that in the asymptotic (long time) regime the net effect of the interaction with the environment
can be represented by the diagram shown in Fig. \ref{fig:interpret}. In the diagram time flows from left to right. 
The original oscillators $x_{1,2}$ are transformed into the virtual oscillators $x_{\pm}$ by the action of an 
ordinary $50/50$ beam splitter. After the beam-splitter the oscillator $x_-$ evolves unitarily and decoupled 
from $x_+$ which, in turn, interacts with the environment. In the long time limit the interaction with the
environment leads to an equilibrium state for $x_+$ which is completely uncorrelated with the state of $x_-$. Such state is
Gaussian and fully characterized by the equilibrium variances $\Delta x_+$ and $\Delta p_+$.  Finally, the second 
beam splitter recombines the two virtual oscillators to produce again the real modes $x_1$ and $x_2$. 

The entanglement in the final state is a quantum resource that can be originated from  other quantum resources available 
in the state of the $x_\pm$ oscillators immediately before the second beam splitter. Indeed, such quantum resource is squeezing. 
In fact, it is well known that a beam splitter can produce entangled states of the outgoing modes provided the input modes 
are squeezed \cite{Kim02}. This observation enables us to understand the origin of the entanglement in the final state 
of the $x_{1,2}$ oscillators: It comes from the squeezing which is available either in the $x_-$ or in the $x_+$ oscillators. 
In turn, the squeezing in the $x_-$ mode, which naturally oscillates with a frequency $\omega_-$, is itself inherited from 
the squeezing (or entanglement) eventually present in the input modes. On the other hand, squeezing in the asymptotic 
(equilibrium) state of the $x_+$ oscillator, which is measured by $r_{crit}$, is also a source for entanglement in the final state.
Below, we will show that there are some situations in which a non-vanishing value for $r_{crit}$ arises as a non-trivial 
(non-Markovian) effect of the environment. 

The three phases we mentioned above can be understood using this interpretation. To make more evident the connection between 
final entanglement and squeezing it is convenient to rewrite the equations for the asymptotic entanglement (\ref{eq:Eoft}) as follows:
\beqa
E(t)&=&|r_{crit}|-S_{crit}+|r| G(t), \ \ \ \ {\rm if}\ |r|\leq |r_{crit}|,\nonumber\\
E(t)&=&|r|-S_{crit}+|r_{crit}| G(t), \ \ \ \ {\rm if}\ |r| > |r_{crit}|.\nonumber
\eeqa
We can extract some interesting conclusion from these equations. First, it is clear that for initial values 
$|r|\le |r_{crit}|$ it is possible to use the environment as a resource from which we extract entanglement.
Thus, for such low squeezing the entanglement in the final state can be larger than the available quantum 
resource (squeezing) present in the initial state. This is indeed the case if the inequality 
$|r_{crit}|-S_{crit}\ge 2|r|$ holds. In other cases the environment degrades the quantum 
resource which is already present in the initial state (either in the form of squeezing 
or entanglement).

With this interpretation, and using the ideas discussed in \cite{Kim02}, 
we can conclude that non-classicality at the output fields (after the second beam splitter)
must arise from some form of non-classicality at the input. This can exist if the equilibrium 
state has some degree of squeezing (which is the case for position coupling) or if the initial state of $x_-$ is non-classical. 
For instance, with initial coherent states the condition for the existence of entanglement in the 
final state ($r=0$, $\delta x_+ \delta p_+ =\delta x_- \delta p_- =1/2$) is $r_{crit}>1/2 \ln(2 \Delta x_+ \Delta p_+)$. 
Thus, to fulfill this condition we  need the environment to produce an equilibrium state where the variance of one 
of its quadratures is smaller than the vacuum limit,  i.e. $\min\{\Delta^2 x_+ , \Delta^2 p_+\}<1/2$  (for $m=1,\Omega_-=1$).

\section{Models for the environment}
\label{sec:model}

In this Section we will discuss in some details two widely used models for system-environment interaction and 
will discuss the nature of the asymptotic state that is obtained. This, as mentioned above, will determine
the way in which entanglement evolves. We consider a system formed by two identical oscillators, with mass $m$
and bare frequency $\omega_0$, whose Hamiltonian is
\beqa
\tilde H_\sys&=&\frac{1}{2m}(p_1^2+p_2^2)
+\frac{m}{2}\omega_0^2(x_1^2+x_2^2)+\nonumber\\
&+&\left(m c_{12}x_1 x_2 + \tilde c_{12}\frac{p_1 p_2}{m \omega_0^2}\right).
\eeqa
The last two terms include a general type of coupling between the oscillators being $c_{12}$ and $\tilde c_{12}$
the corresponding bare coupling constants. The environment will consist of a set of harmonic oscillators whose 
coordinates are labeled as $q_n$. We will consider the coupling between the system and the environment as described by 
\beq
\tilde H_{\sys\env}=x_+ \sum_{n=1}^{N} c_n q_n+
\left(\frac{p_+}{m\omega_0}\right)
\sum_{n=1}^{N} {\tilde c_n\over m_n w_n} \pi_n.
\eeq
Two cases will be discussed in detail. a) Position coupling: This corresponds to the case where $\tilde c_n=0$,
i.e. the coupling is bilinear in the coordinates $x_+$ and $q_n$ (in this case we will consider that the 
coupling between the oscillators is only through position, i.e. $\tilde c_{12}=0$). b) Symmetric coupling:
This corresponds to the case $c_n=\tilde c_n$. In this case the interaction is symmetric under interchange
of position and momentum and can be easily written in terms of creation and annihilation operators. 
It is precisely of the form obtained in the so-called RWA (in this case we will consider that the 
interaction between the oscillators is also symmetric, i.e. $c_{12}=\tilde c_{12}$).

For these two cases, the effect of the environment is determined by the spectral density defined  
as \mbox{$J(w)=\sum_n c_n^2\delta(w-w_n)/2m_n  w_n$}. We will present results for the so-called Ohmic 
environment where $J(w)={2\over \pi}m\gamma_0w \theta(\Lambda-w)$, where $\Lambda$ is a high frequency 
cutoff and $\gamma_0$ is a coupling constant (other spectral densities were analyzed in \cite{PazRoncaglia09}). 
We also assume factorized initial conditions between system and environment,
and that the initial state of the environment is thermal (with temperature $T$).
For the two types of coupling we consider it is possible to obtain an exact master equation
that governs the evolution of the reduced density matrix of the two oscillators. We will briefly describe them now. 

For position coupling the exact master equation was obtained some time ago and reads \cite{HuPazZha92}:
\beqa
\dot\rho&=&-i[H_R,\rho]-i\gamma(t)[x_+,\{p_+,\rho\}]-\nonumber\\
&-&D(t)[x_+,[x_+,\rho]]-f(t)[x_+,[p_+,\rho]].\label{eq:mastereq}
\eeqa
Here, the renormalized Hamiltonian is $H_R=H_\sys+m \delta\omega^2(t)x_+^2/2$. The effect of the environment 
shows up in four terms: The environment induces a renormalization of the frequency of the $x_+$ oscillator, a dissipative
term with a time dependent damping rate $\gamma(t)$ and two diffusive terms with time dependent coefficients $D(t)$ and $f(t)$.
All coefficients depend on the environmental spectral density in a rather complex way ($D(t)$ and $f(t)$ also depend on
the initial temperature $T$). The explicit form of these coefficients was studied elsewhere \cite{HuPazZha92,Fleming07}. 
Here, we will only use the fact that for the Ohmic environment all coefficients approach constant asymptotic values in
the long time limit. 
It is worth pointing out a technical detail concerning the renormalization induced by the environment:
The bare frequencies of the virtual oscillators are $\omega_{\pm }^2=\omega_0^2\pm c_{12}$. The interaction with the environment
renormalizes the frequency of $x_+$ that is shifted according to $\Omega^2(t)=\omega_0^2+c_{12}+\delta\omega^2(t)$
(noticeably, the shift $\delta\omega^2$ diverges in the limit of large cutoff $\Lambda$), while the $x_-$ oscillator
evolves freely with frequency $\omega_{-}^2=\omega_0^2- c_{12}$. From these expressions we can also write 
the renormalized frequency and coupling of the real oscillators
$\Omega_R^2(t)= \omega_0^2 + \delta\omega^2(t)/2$ and $C_{12}(t)=c_{12} + \delta\omega^2(t)/2$.
Therefore for the physical frequencies of both virtual oscillators to be independent of the cutoff one needs 
to absorbe $\delta\omega^2$ in a
renormalization of the bare frequency $\omega_0$ and the bare coupling constant $c_{12}$. 
In such case, the long time oscillations of entanglement become cutoff-independent. 

The master equation can be used to obtain expressions for the variances of position and momentum for the $x_+$ oscillator. 
In fact, for $\Delta^2 x_+=\langle x_+^2\rangle$ and $\Delta^2 p_+=\langle p^2_+\rangle$ we find 
\beq
\Delta p_+= \sqrt{{D \over 2 \gamma}}, \quad \Omega \Delta x_+= \sqrt{{D \over 2 m^2\gamma} - 
{f \over m}}, \label{eq:Dandf}
\eeq
and $\mean{\{x_+,p_+\}}=0$. Where we used upper case letters for the renormalized quantities and we omit the time 
label when referring to value of the coefficient in the long time regime. 

For symmetric coupling an exact master equation exists (for all spectral densities and initial temperatures of the environment). 
In the Appendix we show a simple derivation of such equation valid for arbitrary temperatures. Not surprisingly, the equation 
is nothing but a symmetrized version of the previous one (\ref{eq:mastereq}):
\beqa
\dot\rho&=&-i[\tilde H_R,\rho]-i\tilde \gamma(t)\Big([x_+,\{p_+,\rho\}]-[p_+,\{x_+,\rho\}]\Big)\nonumber\\
&-&\tilde D(t)\Big([x_+,[x_+,\rho]]+ {1\over m_+^2 \omega_+^2} [p_+,[p_+,\rho]]\Big).\label{eq:mastereq2}
\eeqa
Here, the renormalized Hamiltonian is 
$\tilde H_R=\tilde H_\sys+ \delta\tilde\Omega^2(t)\Big({p_+^2\over{2m_+}}+\frac{m_+}{2} \omega_+^2x_+^2\Big)/ \omega^2_+$. 
In this case the effect of the environment is contained in three terms. The renormalization is also symmetric under 
position and momentum interchange. Renormalized mass and frequency of the $x_+$ oscillator must be defined as 
$M(t)=m/(1+(\delta\tilde\Omega^2(t)+ c_{12})/\omega_0^2)$ and   $\Omega(t)=\omega_0(1+(\delta\tilde\Omega^2(t)+c_{12})/\omega_0^2)$.
As expected, both damping and diffusion terms are symmetric under canonical interchange of position and momentum. 
As in the previous case, all the coefficients of the master equation approach constant asymptotic values. As before, 
it is straightforward to obtain the values of position and momentum dispersions: 
\beq
\Delta p_+=M\Omega\Delta x_+= \sqrt{{\tilde{D} \over 2 \tilde\gamma}}; \label{eq:RWADxDp}
\eeq
and $\mean{\{x_+,p_+ \}}=0$. One can notice that, contrary to what happened with the previous model,
the asymptotic state of the oscillator $x_+$ has balanced variances.

\begin{figure}[htb!]
\centering
\leavevmode
\includegraphics[width=8.7cm]{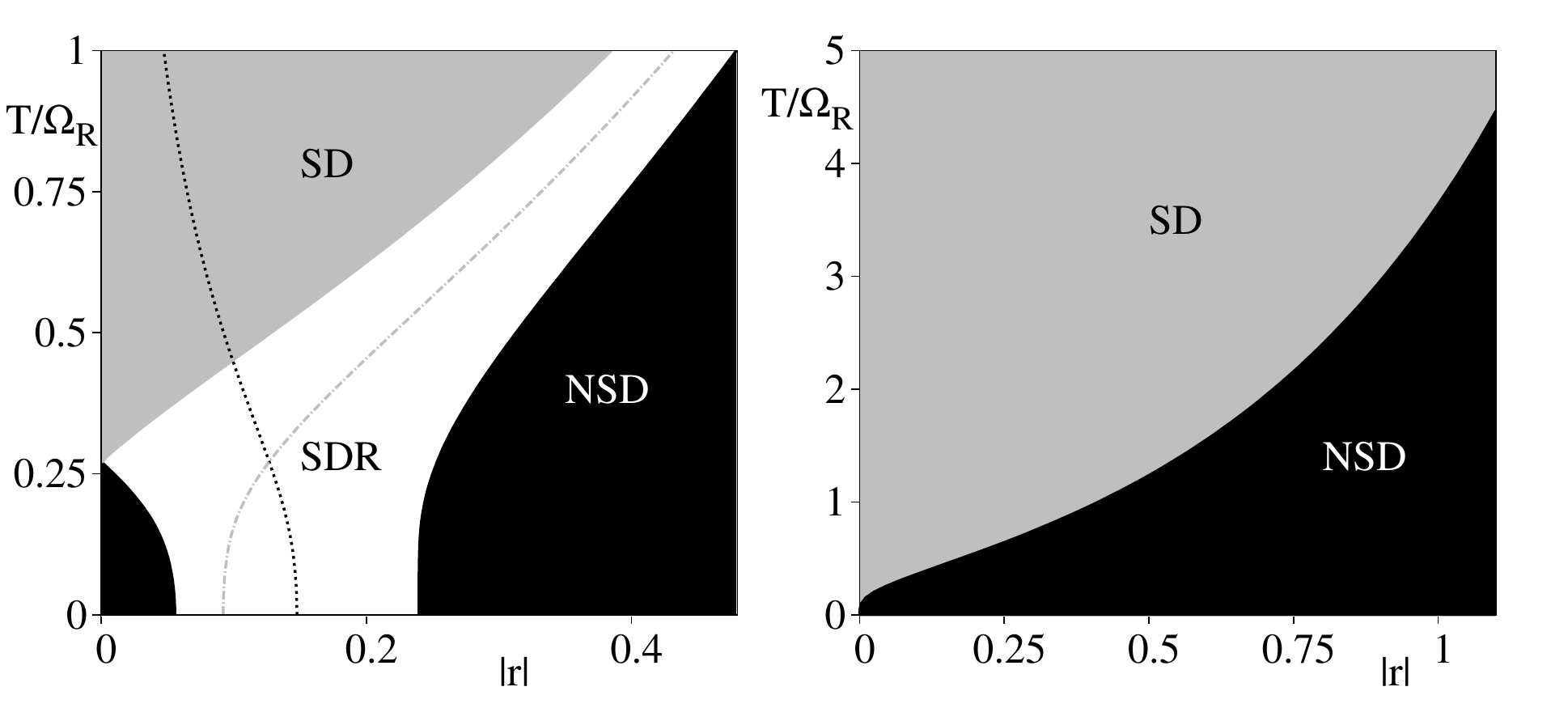}
\caption{Dynamical phases for the entanglement between two non-interacting oscillators, $C_{12}=0$, that are coupled to
the same environment. $r_{crit}$ (black line) and $S_{crit}$ (grey line). 
For position coupling (left panel) there are three phases (NSD, SDR, SD) while for symmetric coupling 
(right panel) only two phases exist. The initial state of the two oscillators is such that the virtual oscillator
$x_-$ is in a pure state with squeezing $r$. The environment is initially in a thermal state with temperature $T$. 
}
\label{fig:old-phasediagram}
\end{figure}

In \cite{PazRoncaglia09} the evolution of entanglement was studied for these two models under the assumption that
the oscillators did not interact directly (i.e. $C_{12}=0$) and initial conditions such that the state of the $x_-$ 
oscillator was pure (i.e. $\delta x_-\delta p_-=1/2$). 
In such case the values of $r_{crit}$ and $S_{crit}$ were obtained for a variety of spectral densities. 
The phase diagrams describing the evolution of entanglement for the two types of coupling were presented and are 
shown in Figure \ref{fig:old-phasediagram}. For position coupling the fact that $r_{crit}$ may be non-vanishing opens 
the door to the existence of an NSD phase for low temperatures and small squeezings. It is also responsible for the
existence of the SDR phase. These two effects disappear for the case of symmetric coupling, that only exhibits two phases. 
Below, we will generalize these results. It is also worth pointing out that the above results are valid for 
resonant oscillators with equal coupling to the environment, for non-resonant oscillators \cite{PazRoncaglia09} 
or spatially separeted oscillators \cite{Zell09}, the asymptotic entanglement becomes independent of the initial 
state and sudden death occurs above a critical temperature or distance.

\section{Generalizations}
\label{sec:generalizations}

\subsection{Initially mixed states}

\begin{figure}[htb!]
\centering
\leavevmode
\includegraphics[width=8.7cm]{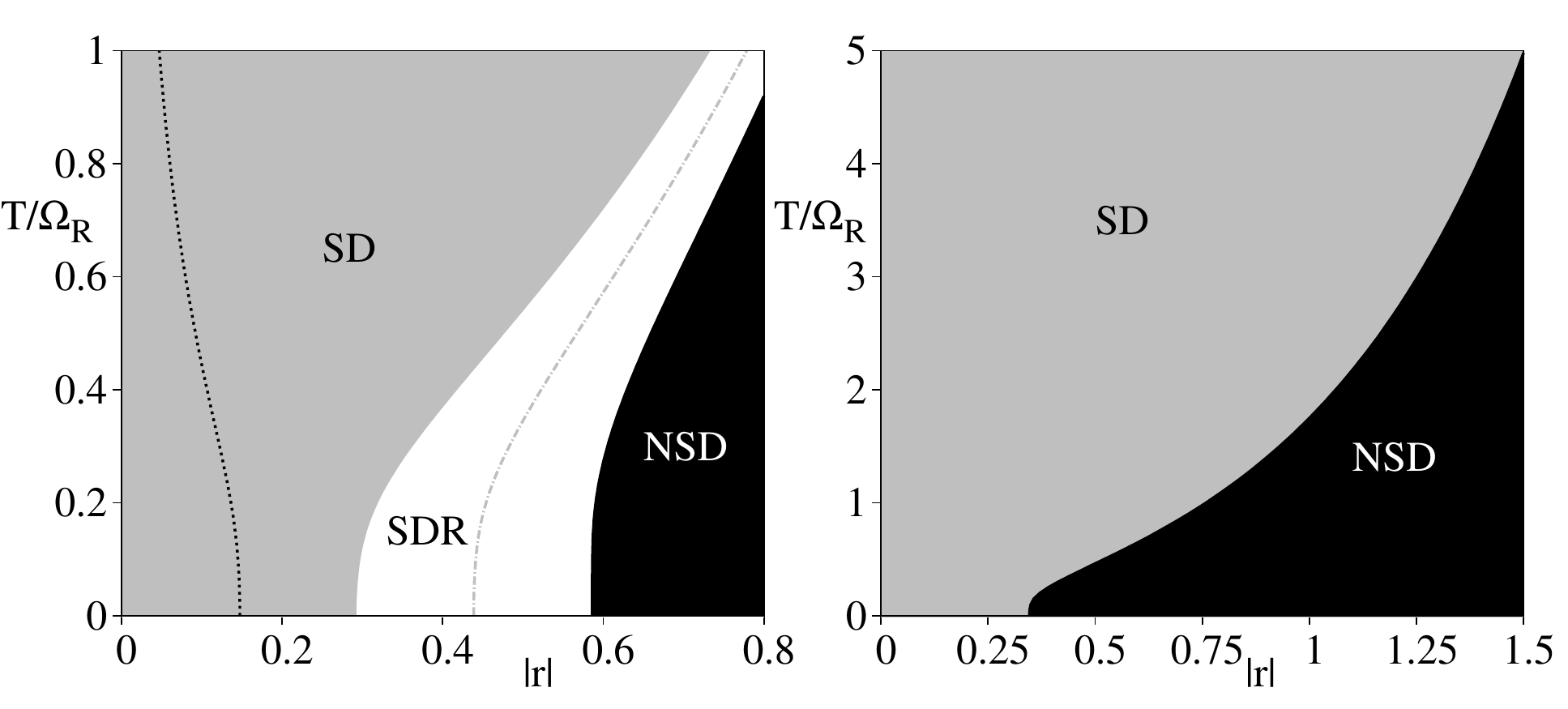}
\caption{Dynamical phases for the entanglement between two non-interacting oscillators 
 initially in a state such that the virtual oscillator $x_-$ is in a mixed state $\delta x_-\delta p_-=1$,
for position coupling (left panel) and symmetric coupling (right panel). 
}
\label{fig:phase-mixed}
\end{figure}

As a first generalization we consider the case where the initial state of the virtual oscillator $x_-$ is mixed.
It is clear that for this situation all the above formulas apply. The only footprint of the initial state of the system
appears through the dependence of  $S_{crit}$ on the initial dispersions of the $x_-$ oscillator. The purity of the 
state of $x_-$ is characterized by the product $\delta x_- \delta p_-$ that appears in $S_{crit}$. When this product 
is increased the value about which the entanglement oscillates decreases (as can be seen from the equations 
(\ref{eq:ent}) and (\ref{eq:Scrit})). This implies that as a consequence of the impurity present in the $x_-$ 
oscillator the final entanglement decreases. This situation applies, for instance, to the case in which the initial 
state of each real oscillator is mixed or (using the quantum optical analogy) for an initial pure state that ends 
entangled after the application of the first beam-splitter. The phase diagram for these states can be obtained 
in a simple way from the diagram for initial pure states by a translation of the curve $S_{crit}$ to the right. 
The net effect of this change is to move upwards the horizontal axis.  As a consequence, the value of the critical 
temperature below which the NSD phase exists becomes lower. In fact, the NSD island  can disappear depending on the 
degree of purity of the state of $x_-$. The phase diagram for both types of couplings is shown in Figure \ref{fig:phase-mixed}.

\subsection{Interacting oscillators}

\begin{figure}[htb!]
\centering
\leavevmode
\includegraphics[width=8.7cm]{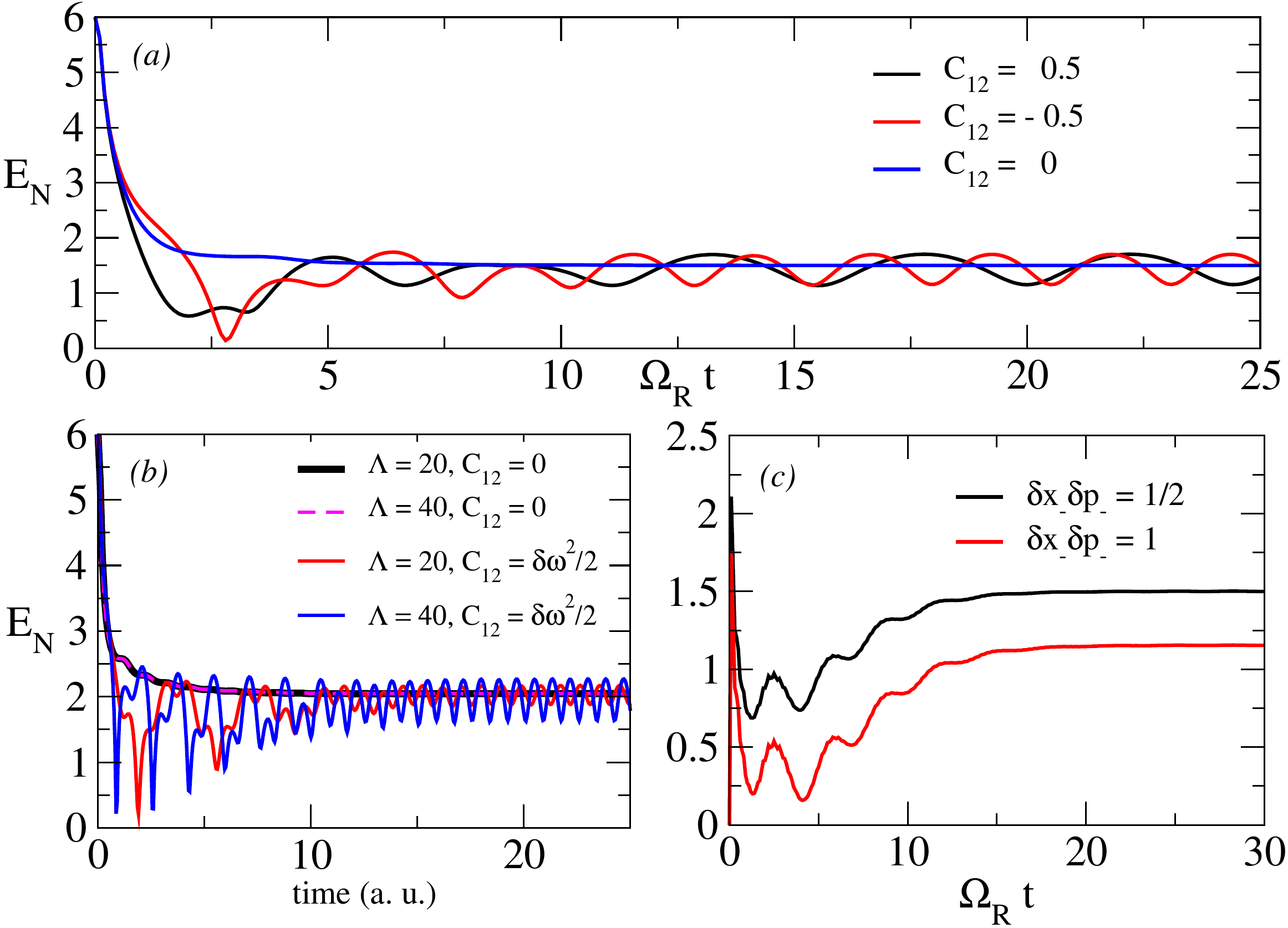}
\caption{Dynamics of the entanglement for resonant oscillators and position coupling at $T/\Omega_R=10$, $\gamma_0=0.1$ and $\Lambda=20$.
$(a)$ Initial two-mode squeezed state ($r=3$, $\Omega_{1,2}\equiv\Omega_R=1$) with different couplings,
we observe oscillations at high temperatures when there is non-vanishing renormalized coupling $C_{12}$.
$(b)$ Initial two-mode squeezed state ($r=3$, $\Omega_R=3$), oscillations that
depend on the cutoff frequency are present when one consider a vanishing bare coupling (i.e. $c_{12}=0$ and $C_{12}=\delta \omega^2/2$).
$(c)$ Separable initial state ($r=3$), $C_{12}=0$. The final entanglement depends on the 
degree of purity of the initial state and is lower than the entanglement achieved
for an initial pure state in a quantity given by
$\ln\left[2 \delta x_-\delta p_- \right]/2$.
}
\label{fig:margPlot}
\end{figure}

If the renormalized coupling $C_{12}=c_{12}+{\delta\omega^2/2}$ is non-vanishing the analysis is also a straightforward 
extension of the previous one. When the coupling with the environment is through position, the renormalized frequencies
of the oscillators $x_\pm$ are $\Omega_\pm^2=\Omega_R^2\pm C_{12}$, 
where $\Omega_R$  is the renormalized frequency of the $x_{1,2}$ oscillators. 
The non-vanishing coupling induces different frequencies for both virtual oscillators. In such case the value of $r_{crit}$ is:
\beq
r_{crit}={1\over 2}\ln\left[m \Omega {\Delta x_+\over\Delta p_+}\right]+{1\over 4}\ln\left[\omega_-\over \Omega\right].
\eeq
The most important difference with the previous case of non-interacting oscillators is that $r_{crit}$ can be 
different from zero even in the limit where the $x_+$ oscillator is not squeezed. As a consequence, it is possible
to observe oscillations of the entanglement in the high temperature regime. Some examples of the behavior of the entanglement 
in this situation are shown in Fig. \ref{fig:margPlot} $(a)$. It is interesting to stress that, as we mentioned above, in all cases 
it is necessary to include a bare coupling term proportional to $\delta \omega^2$ in the Hamiltonian. 
Thus, only in this way the system would not oscillate with a cutoff-dependent frequency in the long time limit 
(see Fig. \ref{fig:margPlot} $(b)$ and $(c)$). 

\begin{figure}[htb!]
\centering
\leavevmode
\includegraphics[width=8.7cm]{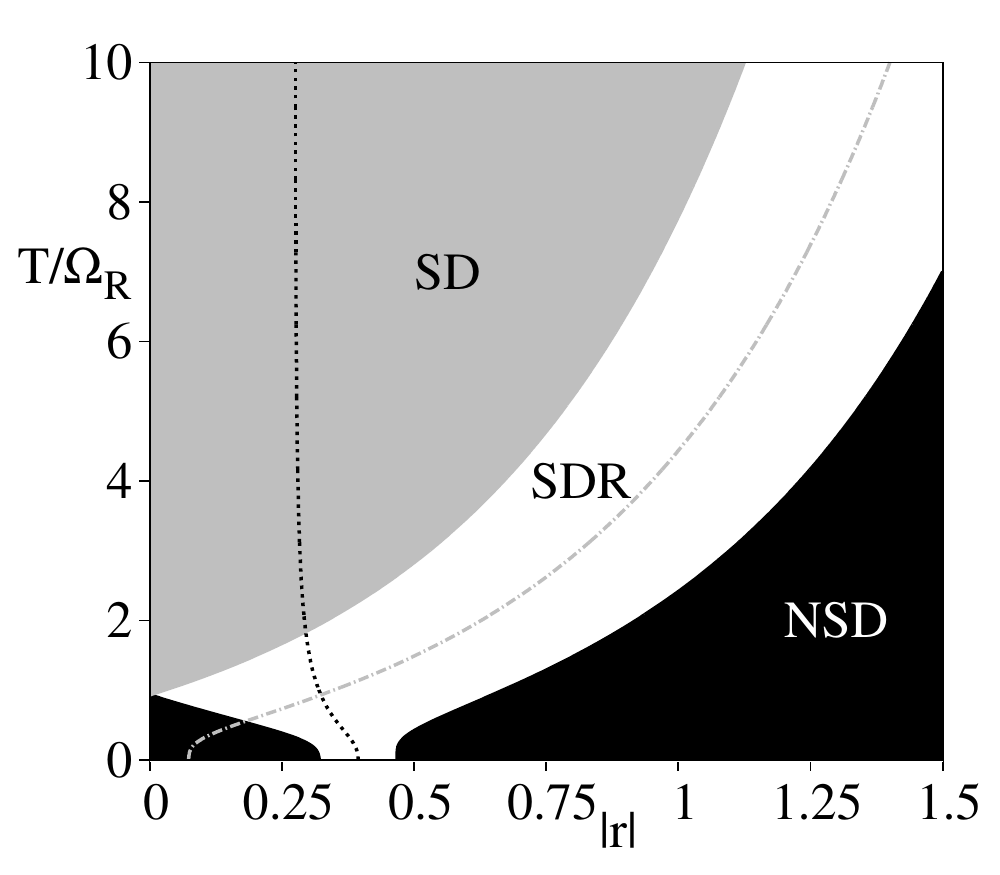}
\caption{Dynamical phases for the entanglement between two interacting oscillators, with $C_{12}=-0.5$, 
that are coupled to the same environment. The initial state is such that
$\delta x_- \delta p_-=1/2$. $r_{crit}$ achieves a 
constant value at high temperatures, and a SDR region appears at high temperatures.
}
\label{fig:phase-interact}
\end{figure}

For the case of symmetric coupling the asymptotic behavior does not change
considerably if we add an interaction between the oscillators (provided the interaction is also symmetric in position and momentum). 
In fact, in this case we always obtain $M\Omega=m_-\omega_-=m\omega_0$ and a vanishing $r_{crit}$, 
as it can be seen from the equations (\ref{eq:rcrit}) and (\ref{eq:RWADxDp}). 
As a consequence, for the symmetric coupling, entanglement
is constant in the asymptotic regime and the phase diagram does not change. The phase diagram for position coupling with 
interacting oscillators is shown in Figure \ref{fig:phase-interact}.

\section{Conclusions}
\label{sec:conc}
In summary, we have presented a simple general overview of the evolution for the entanglement between two oscillators 
coupled to a common Ohmic reservoir. We showed how the existence of non-trivial phases for the evolution of the entanglement 
can be seen by using a simple interpretation based on a quantum optical analogy. The fact that the $x_\pm$ virtual oscillators 
decouple and that the $x_+$ oscillator approaches equilibrium is what makes this interpretation possible. 
The nature of the equilibrium state for $x_+$ may be peculiar since it may be squeezed by a factor $r_{crit}$ when
the coupling with the environment is not symmetric under position-momentum interchange. In the final Section, we generalized 
the results previously obtained in \cite{PazRoncaglia08,PazRoncaglia09} to include cases where the state of the virtual oscillators
is mixed as well as to consider the case of interacting systems. In all cases, the evolution of entanglement can be discussed 
in terms of a phase diagram that describes all qualitatively different behaviors in the long time limit. It is worth noticing 
that the phase diagram for interacting oscillators shown in Figure \ref{fig:phase-interact}, that includes the three dynamical
phases (NSD, SDR and SD), seems to be observable in experiments realizable with current technologies in ion traps \cite{CormickPaz09}.

\acknowledgements
JPP is a member of CONICET and AR has a fellowship from CONICET. This work was supported with grants from ANPCyT (Argentina) 
and Santa Fe Institute (SFI, USA).

\appendix
\section{Exact master equation for symmetric coupling in position and momentum}
\label{ap:ME}

We present here a brief sketch of a simple derivation of the exact master equation for symmetric coupling which
is similar to the one obtained for position coupling in \cite{Halliwell96}. The derivation is valid for all spectral
densities and temperatures. The full Hamiltonian, written in terms of creation and annihilation operators is:
\beq
H=\hbar \omega a^\dagger a+\sum_k\hbar w_k b_k^\dagger b_k+
\sum_k g_k (a b_k^\dagger+b a_k^\dagger).
\eeq
The first step of the derivation is to notice that this Hamiltonian preserves the Gaussian nature of the states 
(this is obvious since $H$ is a quadratic form of the coordinates and momenta of all the particles). Therefore,
the evolution operator of the reduced density matrix of the system is a Gaussian operator also. It is possible to show,
following the same steps described in the derivation contained in \cite{PazZurekLesHouches}, that if the propagator is Gaussian the
master equation is local in time. Moreover, it has a limited number of terms whose number is further reduced if one 
imposes the constraint that the equation should be symmetric under canonical exchange of position and momentum. 
Thus, under this condition one can show that the form of the master equation should be
\beqa
 \dot\rho&=&-i[\tilde H_R(t),\rho]+\left({\tilde D(t)\over m\omega}+\tilde\gamma(t)\right)\left(2a\rho a^\dagger
-a^\dagger a \rho-\rho a^\dagger a\right)\nonumber \\ 
&&+\left({\tilde D(t)\over m\omega}-\tilde\gamma(t)\right)\left(2a^\dagger \rho a
-a a^\dagger\rho-\rho a a^\dagger\right).
\label{eq:MERWA}
\eeqa
This equation contains three unknown coefficients with a clear physical interpretation: A renormalized frequency
in $H_R(t)$ such that $\Omega_R(t)=\omega+\delta\tilde\Omega^2(t)/\omega^2$, a dissipation coefficient $\tilde\gamma(t)$
and a diffusion coefficient $\tilde D(t)$. The above argument simply tells us that the master equation should have this 
form but does not enforce any constraint in the time dependence of such coefficients. Now, we will find them using
the following argument. 

From the total Hamiltonian, we can derive Heisenberg equations for all the operators, which turn out to be:
\beqa
{da\over dt}&=&i\omega a-i\sum_k g_k b_k, \label{eq:mot-a} \\
{db_k\over dt}&=&iw_k b_k-i g_k a, \label{eq:mot-bk}\\
{d(aa^\dagger+a^\dagger a)\over dt}&=&2 i\sum_k g_k(a b_k^\dagger-b_k a^\dagger).
\label{eq:mot-aa}
\eeqa
These can be formally solved as
\beqa
&&a(t)= ua(0)+\sum_n p_n b_n(0),\label{eq:sola}\\
&&b_k(t)=d_k a(t)+\sum_n q_{kn} b_n(0),
\label{eq:solb}
\eeqa
where $u$, $p_n$, $d_k$ and $q_{kn}$ are appropriate time-dependent coefficients. 

On the other hand, the master equation (\ref{eq:MERWA}) can be used to obtain evolution equation for
expectation values of the operators of the system which turn out to be:
\beqa
&&{d\mean a\over dt}=-\left(2\gamma(t)+i\Omega_R(t)\right)\mean a, \\
&&{d\mean{aa^\dagger+a^\dagger a}\over dt}=-4\gamma(t)\mean{aa^\dagger+a^\dagger a}+4{\tilde D(t)\over m\omega}.
\eeqa
Comparing equations (\ref{eq:mot-a}) with the expectation value of equation  (\ref{eq:mot-aa}) we can simply
obtain the expressions for the time dependent coefficients. Moreover, imposing that the initial state of 
the environment is thermal, $\mean{b_k^\dagger(0) b_k(0)}=n_k$, these coefficients can be expressed as:
\beqa
\tilde\gamma(t)&=&{i\over 4}\sum_k g_k(d_k-d^*_k),\nonumber\\ 
 {\delta\tilde\Omega^2(t)\over \omega^2}&=&{1\over 2}\sum_k g_k(d_k+d^*_k),\\
{\tilde D(t)\over m \omega}&=&{i\over 4}\sum_{k,l}g_k(q_{kl}^*p_l-q_{kl}p_l^*)(2n_l+1). \nonumber
\eeqa
Hence, they are completely defined in terms of the solution to the equation
of motion (\ref{eq:sola})-(\ref{eq:solb}). 

The solutions above equations can be explicitly written. In fact, from (\ref{eq:mot-bk}) we can write:
\beq
b_k(t)=e^{-iw_kt}b_k(0)-ig_k \int_0^t e^{-iw_k(t-s)} a(s) ds.
\label{eq:sol-bk}
\eeq
Moreover using (\ref{eq:mot-a}) and (\ref{eq:sol-bk}) we get
\beq
{da(s)\over ds}+i\omega a(s)+\int_0^s \eta(s-x)a(x)dx=if(s),
\label{eq:at}
\eeq
where $f(s)=-\sum_k g_k e^{-iw_ks} b_k(0)$; and the kernel $\eta(s)$ is defined as
\beq
\eta(s)=\int_0^\infty dw J(w) e^{-iw s}.
\label{eq:kernel}
\eeq
Here, the spectral density is $J(w)=\sum_k g_k^2 \delta(w-w_k)$. 
Now, the equation (\ref{eq:at}) can be solved using the Laplace transform and the convolution theorem, with
initial and final conditions given by: $a(s=0)=a(0)$ and $a(s=t)=a(t)$. From these expressions it would be possible 
to find the time-dependence of all the coefficients of equations (\ref{eq:sola})-(\ref{eq:solb}).
Finally, we note that there are certain relations satisfied by the coefficients: One of them arises from the fact that the 
commutation relations are preserved during the evolution. Thus, as $[a(t),b_k(t)^\dagger]=0$, one has 
\mbox{$d_k=-\sum_n q_{kn} p_n^*$}. This relation can be used to show that 
at zero temperature we have $\tilde D(t)/m\omega=\tilde \gamma(t)$. And the exact
master equation has only two time-dependent coefficients: frequency renormalization and dissipation.

\end{document}